\documentclass[twocolumn,aps,prl] {revtex4} 
\usepackage{graphicx}
\begin{document}
\pagestyle{empty} 
\title{Contact mechanics with adhesion: Interfacial separation and contact area}
\author{C. Yang$^1$,  B.N.J. Persson$^1$, J. Israelachvili$^2$ and K. Rosenberg$^2$}
\affiliation{$^1$Institut f\"ur Festk\"operForschung, Forschungszentrum J\"ulich, D-52425 J\"ulich, Germany}
\affiliation{$^2$Department of Chemical Engineering, University of California, Santa Barbara, California 93106, USA}

\begin{abstract}
We study the adhesive contact between elastic solids with randomly rough, self affine fractal 
surfaces. We present molecular dynamics (MD) simulation results for the interfacial
stress distribution and the wall-wall separation. We compare the MD results for
the relative contact area and the average interfacial separation, with the prediction of the
contact mechanics theory of Persson. We find good agreement between theory and the simulation results. 
We apply the
theory to the system studied by Benz et al.\cite{MB} involving polymer in contact with
polymer, but in this case 
the adhesion gives only a small modification of the interfacial separation 
as a function of the squeezing pressure.  

\end{abstract}
\maketitle


With the rapid development of micro/nano electromechanical devices in last decade,
surface forces play a more and more important role in modern technology. This is due to the increase of 
the ratio between the number of atoms on the surface and that in the volume.  
When we bring two surfaces together, attractive (and repulsive) forces act between them,
and a non-zero force is often required to separate two solid bodies
placed in intimate contact\cite{JKRP,Fuller}, a phenomenon referred to as adhesion.

Adhesion manifests itself in many ways. Thus adhesion on one hand makes it
possible for a Gecko to walk on the ceilings or run on a vertical wall\cite{Gao,Autumn}. 
On the other hand, adhesion can lead to the failure of micro or nano devices, 
e.g. micro-sized cantilever beams\cite{Adhesion}. Thus, if it is too long or too thin, the 
free energy minimum state corresponds to
the cantilever beam partly bound to the substrate, which leads to the failure of the device. 
However, if the surface roughness is increased, 
the non-bonded cantilever state may 
be stabilized due to the decrease of cantilever-substrate binding energy.

In reality most surfaces are not atomically flat. 
Even if a surface appears flat at low magnification, when we study the surface at higher magnification
we usually observe surface roughness on small length-scale. Similarly, when two solids with nominally smooth surfaces
are brought into contact, generally they do not make contact everywhere, but at high enough
magnification one usually observes many non-contact regions. 
The study of the interfacial separation is essential for describing, 
e.g., sealing \cite{Sealing}, capillary adhesion \cite{Capillary.adhesion} or optical interference. 

Contact mechanics between solid surfaces is the basis for understanding many
tribology processes\cite{Bowden,Johnson,BookP,Isra} 
such as friction, adhesion, wear and sealing. The two most important
properties in contact mechanics are the area of real contact and the interfacial separation between the
solid surfaces. 
For non-adhesive contact and small squeezing pressure, 
the average interfacial separation depends 
logarithmically on the squeezing pressure\cite{MB,P4},
and the (projected) contact area depends linearly on the squeezing pressure\cite{P1,Greenw,Muser}.
For adhesive contact, however, no numerical results have been presented in the literatures to 
test the contact mechanics theory with adhesion.
In this letter, we study the relation between 
the average interfacial separation and the squeezing pressure when
adhesion is included. We compare the result of MD-simulations with 
a recently developed contact mechanics theory\cite{P1,P2,P3,P4}. We find good agreement with the theory,
which represents the first test of the theory when the adhesive interaction 
is included in the analysis.

We review the contact mechanics theory of Persson briefly.
It can be used to calculate the
stress distribution at the interface, the area of real contact and the average interfacial
separation between the solid walls\cite{P1,P4}. In this theory, the interface is studied at different
magnifications $\zeta = L/\lambda$ where $L$ is the linear size of the system and $\lambda$
the resolution. The wavevectors are defined as $q=2 \pi /\lambda$ and $q_L=2 \pi /L$ so that
$\zeta=q/q_L$.

Consider an elastic block with a flat
surface in adhesive contact with a hard substrate with a randomly rough surface.
Let $\sigma({\bf x},\zeta)$ denote the (fluctuating) stress at the interface between the solids when
the system is studied at the magnification $\zeta$. The distribution of interfacial stress
\begin{equation}
\label{equ1}
P(\sigma,\zeta) = \langle \delta (\sigma- \sigma({\bf x},\zeta))\rangle.
\end{equation}
In this definition we do not include the $\delta (\sigma )$-contribution from 
the non-contact area.

For perfect (or complete) contact it is easy to show that $P(\sigma,\zeta )$ satisfies\cite{P1}
\begin{equation}
\label{equ2}
{\partial P \over \partial \zeta} = f(\zeta) {\partial^2 P\over \partial \sigma^2}, 
\end{equation}
where 
$$f(\zeta ) = {\pi \over 4} {E^*}^2 q_L q^3 C(q).$$
Here $E^* = E /(1-\nu^2)$ is the effective elastic modulus.
The surface roughness power spectrum
$$C(q) = {1\over (2\pi )^2}\int d^2x  \ \langle h({\bf x})h({\bf 0})\rangle e^{-i{\bf q}\cdot {\bf x}}$$
where $z=h({\bf x})$ is the surface height at the point ${\bf x}=(x,y)$ and where $\langle .. \rangle$
stands for ensemble average. The basic idea is now to assume that (\ref{equ2})
holds locally also for incomplete
contact. 

To solve (\ref{equ2}) one needs boundary conditions. If we assume that, 
when studying the system at the lowest magnification $\zeta=1$ 
(where no surface roughness can be observed, i.e.,
the surfaces appear perfectly smooth), 
the stress at the interface is constant and equal to $p=F_{\rm N}/A_0$,
where $F_{\rm N}$ is the load and $A_0$ the nominal contact area, then 
$P(\sigma, 1) = \delta (\sigma -p)$.
In addition to this ``initial condition'' we need two boundary conditions along
the $\sigma$-axis. Since there can be no infinitely large stress at the interface we require
$P(\sigma, \zeta) \rightarrow 0$ as $\sigma \rightarrow \infty$. 
For adhesive contact, which interests us here, tensile stress occurs at the interface close 
to the boundary lines of the contact regions. In this case we have the boundary condition
$P(-\sigma_{\rm a},\zeta)=0$, where $\sigma_{\rm a}>0$ is the largest tensile stress possible.
The detachment stress $\sigma_{\rm a} (\zeta)$ depends on the magnification and can be related to
the effective interfacial energy (per unit area) $\gamma_{\rm eff}(\zeta)$ using
the theory of cracks\cite{P3}
$$\sigma_{\rm a} (\zeta) \approx \left ({\gamma_{\rm eff}(\zeta) E q \over 1-\nu^2}\right )^{1/2},$$
where
$$\gamma_{\rm eff}(\zeta) A^*(\zeta) = \Delta \gamma A^*(\zeta_1)- U_{\rm el} (\zeta),$$
where $A^*(\zeta)$ denotes the total contact area at the magnification $\zeta$, 
which is larger than the projected contact area $A(\zeta)$. $U_{\rm el} (\zeta)$ is the elastic energy stored
at the interface due to the elastic deformation of the solids on length scale shorter than $\lambda = L/\zeta$,
necessary in order to bring the solids into adhesive contact (see below).

From (\ref{equ2}) it follows that 
the area of apparent contact (projected on the $xy$-plane) 
at the magnification $\zeta$, $A(\zeta)$, normalized
by the nominal contact area $A_0$, 
can be obtained from
\begin{equation}
 \label{contact.area}
{A(\zeta)\over A_0} = \int_{-\sigma_{\rm a}(\zeta)}^\infty d\sigma \ P(\sigma, \zeta)
\end{equation}
We denote $A (\zeta) / A_0 =P_p (q)$, where the index $p$ indicates that $A (\zeta) / A_0$
depends on the applied squeezing pressure $p$.
The area of (apparent) contact at the highest magnification $\zeta=\zeta_1$ gives the real contact area.
For the elastic energy $U_{\rm el}$ we use\cite{YP}
\begin{equation}
\label{equ3}
U_{\rm el} \approx A_0 E^* {\pi \over 2} \int_{q_L}^{q_1} dq \ q^2W(q,p)C(q),
\end{equation}
where $q_L$ and $q_1$ are the smallest and the
largest surface roughness wave vectors, and\cite{YP}
$$W(q,p) = P_p(q) \left [\beta +(1-\beta) P_p^2(q)\right ],$$
where $\beta = 0.4$.
The equations given above are solved as described in Ref.~\cite{P2}.

Let us now consider the (average) interfacial separation $\bar u$ as a function of the
squeezing pressure $p=F_{\rm N}/A_0$. Note that as $p$ increases, $\bar u$ decreases
and we can consider $p = p(\bar u)$ as a function of $\bar u$. 
Energy conservation gives\cite{P3}
\begin{equation}
\label{equ4}
\int_{\bar u}^\infty du \ p(u) A_0 =U
\end{equation}
where $U=U_{\rm el}+U_{\rm ad}$ is the sum of the elastic energy $U_{\rm el}$ stored at the interface 
and given by (\ref{equ3}), and the
adhesional energy $U_{\rm ad} = -\Delta \gamma A^*(\zeta_1)$. 
From (\ref{equ4}) we get
$$p(\bar u) = - {1\over A_0}{d U (\bar u) \over d\bar u}$$
We can also consider $U$ as a function of $p$ and write
$$p(\bar u) = - {1\over A_0} {d U \over dp} {dp \over d\bar u}$$
or
$$d \bar u = - {1\over A_0 p} {d U \over dp} dp$$
Integrating from $u=0$ (corresponding to $p=\infty$) to $u$ 
(corresponding to the pressure $p$)
gives
\begin{equation}
 \label{interfacial.separation}
\bar u= {1\over A_0} \int_p^\infty dp \ {1\over p} {dU\over dp}
\end{equation}
which is very convenient for numerical calculations.

Let us provide some details about the numerical simulations.
The molecular dynamics system has lateral dimension $L_{x}=N_{x}a$ and $L_{y}=N_{y}a$, 
where $a$ is the lattice spacing of the block.
In order to accurately study contact mechanics between elastic solids,
it is necessary to consider that the thickness of the block is (at least) of the same
order of the lateral size of the longest wavelength roughness on the substrate.
We have developed a multiscale MD approach to study contact mechanics\cite{YTP}.
Periodic boundary condition has been used in $xy$ plane.
For the block $N_{x}=N_{y}=400$, while the lattice space of the substrate
$b\approx a/\phi$, where $\phi=(1+\sqrt{5})/2$ is the golden mean, in order to 
avoid the formation of commensurate structures at the interface. 
The mass of the block atoms is 197 a.m.u. and the $a=2.6 \ \rm \AA$.
The elastic modulus and Poisson ratio of the block are
$E=77.2 \ {\rm GPa}$ and $\nu=0.42$.
For self-affine fractal surfaces, the power spectrum has power-law behavior $C(q) \sim q^{-2(H+1)}$,
where the Hurst exponent $H$ is related to the fractal dimension $D_f$ of the surface
via $H=3-D_f$. For real surfaces this relation holds only for a finite wave vector region
$q_0<q<q_1$. 
Note that in many cases, there is a roll-off wave vector $q_0$ below which
$C(q)$ is approximately constant. 
Here $q_L=2\pi/L, q_0=3q_L, q_1=12 q_L$. $q_0$ is named as roll-off wave-vector.
The physical meaning is that by choosing $q_0=3q_L$ one can obtain a self-average equivalent to an average over $9$ independent samples. 
In MD simulations, the substrate is rigid and fractal with fractal dimension
$D_f=2.2$ and root-mean-square roughness $h_{\rm rms}=10 \ {\rm \AA}$.  
The calculations are carried out under the temperature $0 \rm K$.

\begin{figure}
\includegraphics[width=0.45\textwidth,angle=0]{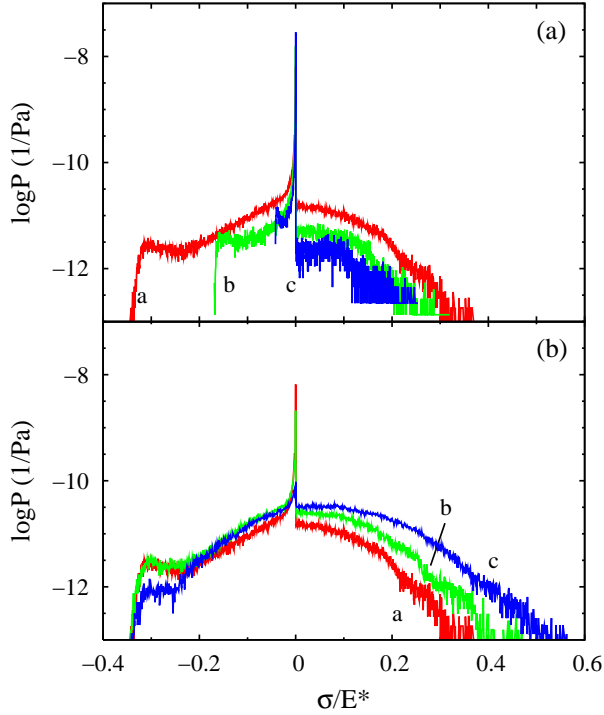}
\caption{\label{multi.pres}
The logarithm (with 10 as basis) of the 
probability distribution of normal stress $\sigma$ (where $\sigma$ is in units of 
$E^{*}$) for (a) 
three different adhesion
parameters $\epsilon=8\epsilon_0$, $4\epsilon_0$ and $\epsilon_0$,
referred to as a, b and c respectively, and (b) with $8\epsilon_0$ for three different pressures
$p/E^{*} = -0.00532$, $0.00832$, $0.06265$ denoted by a,b, and c respectively.} 
\end{figure}

\begin{figure}
\includegraphics[width=0.45\textwidth,angle=0]{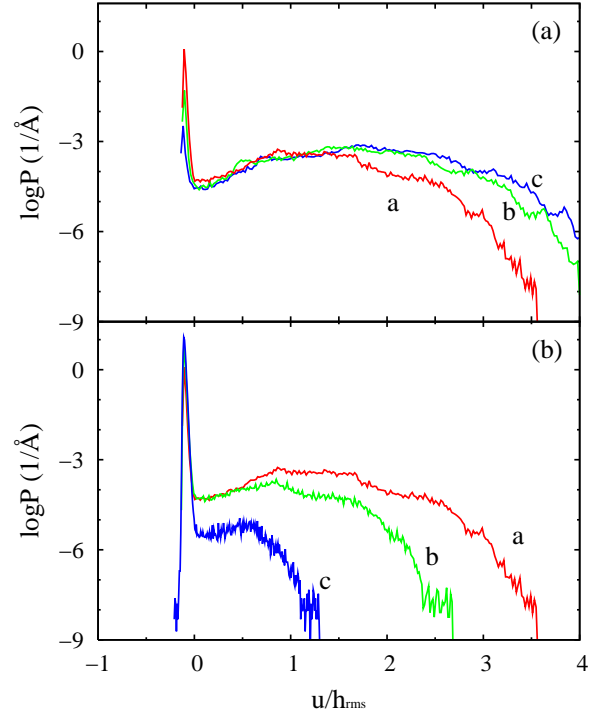}
\caption{\label{multi.sep}
The logarithm (with 10 as basis) of the 
probability distribution of interfacial separations $u$ (where $u$ is in units of the root-mean-square
roughness amplitude $h_{\rm rms}$) for (a) 
three different adhesion
parameters $\epsilon=8\epsilon_0$, $4\epsilon_0$ and $\epsilon_0$,
referred to as a, b and c respectively, and (b) with $8\epsilon_0$ for three different pressures
$p/E^{*} = -0.00532$, $0.00832$, $0.06265$ denoted by a,b, and c respectively.} 
\end{figure}

The atoms at the interface between block and substrate interact with
the potential
$$V(r)=4\epsilon \left[ \left(\frac{r_0}{r} \right)^{12} -\left( \frac{r_0}{r} \right)^{6} \right] $$
where $r$ is the distance between the pair of atoms. The parameter
$\epsilon$ is the binding energy between two atoms at separation $r=2^{1/6}r_{0}$.
In the calculations presented below we have used the $r_{0}=3.28 \ {\rm \AA}$ and
$\epsilon = \epsilon_{0}$, 
$4\epsilon_{0}$ 
and $8\epsilon_{0}$, 
where $\epsilon_0=18.6 \ {\rm meV}$. 
By comparing the total energy for the surfaces separated
with the case where the surfaces are in contact at equilibrium (at zero external load)
we obtain $\Delta \gamma = 0.69 \ {\rm J/m^2}$ for the case $\epsilon = \epsilon_{0}$, and 4 and
8 times higher interfacial binding energy for the other two cases respectively. 

Now, let us discuss how to define contact on the atomic scale when adhesion 
is included. In the absence of adhesion we have found that the interfacial
stress distribution gives the 
most accurate way of deducing the area of real contact\cite{YTP}. When adhesion is included  
we use a cut-off length $d_c$ to define contact. 
It is clear from the force-distance curve that
the only distinctive point is the maximum tensile stress\cite{G1997}. 
Thus the solids are regarded 
as in contact when the stress increases with separation, and separated when the stress decreases
with separation. In our case, the critical wall-wall distance is $d_c=3.68 \ {\rm \AA}$. 
When the interfacial separation $d<d_c$, it is defined as contact. Otherwise it is non-contact. 
More discussions about $d_c$ can be found in Ref.~\cite{Yang.thesis}.

The probability distribution of (perpendicular) stress $\sigma$ 
at the interface is shown in Fig.~\ref{multi.pres}
for (a) the adhesion parameter $\epsilon= \epsilon_0$, $4\epsilon_0$ and $8\epsilon_0$, 
and (b) for $\epsilon = 8 \epsilon_0$ for three different
values of the applied stress $p$ ($p/E^* = -0.00532$, 0.00832 and 0.06265). 
Note that when $\epsilon$ increases, the maximum $\sigma_c$ of the tensile stress 
increases roughly proportional to $\epsilon$. In fact, we expect $\sigma_c \propto \epsilon / a$,
where $a$ is of order a bond length. At the same time the maximum repulsive pressure
increases but weaker than linear. The increase in repulsive stress is, of course, due to the
additionally adhesional load which acts on the block. 

When the applied load or pressure increases, the maximal tensile stress is unchanged,
see Fig.~\ref{multi.pres}(b). This is the expected result since the maximum tensile stress is associated
with breaking the atomic bonds at the edges of the contact regions (which can be considered as crack
tips), and this stress is of course independent of the load. However, the maximum 
repulsive stress increases with load, but the effect is quite small considering the large change in the 
load. 

The probability distribution of interfacial separations is shown in Fig.~\ref{multi.sep}. In
Fig.~\ref{multi.sep}(a) we vary the adhesion parameter $\epsilon$ and in (b) the applied stress
$p$ as in Fig.~\ref{multi.pres}.
When the adhesion increases, the surfaces are pulled closer to each other and the distribution of separations
becomes narrower. Similarly, when the applied pressure increases, the separation between the
walls decreases.  

\begin{figure}
\includegraphics[width=0.45\textwidth,angle=0]{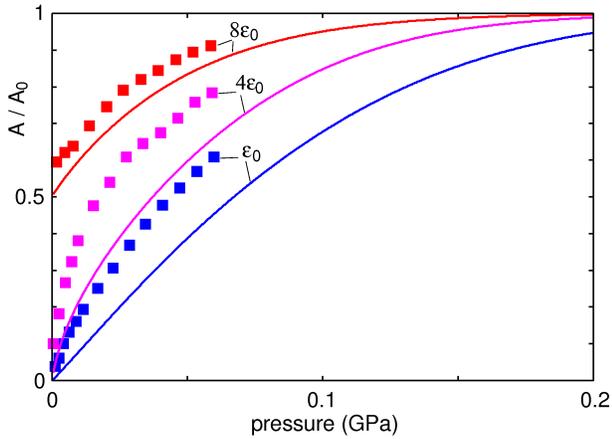}
\caption{\label{last}
Molecular dynamics (square symbols) and theoretical (solid lines) results for the 
relative contact area $A/A_0$ as a function of the squeezing pressure (in GPa)
In the calculation (solid lines) we have used $\Delta \gamma = 0.7$, $2.8$ and $5.6 \ {\rm J/m^2}$
for the curves indicated by $\epsilon_0$, $4 \epsilon_0$ and $8 \epsilon_0$,
respectively. The theoretical results are obtained from Eq.~\ref{contact.area}.
}
\end{figure}

\begin{figure}
\includegraphics[width=0.45\textwidth,angle=0]{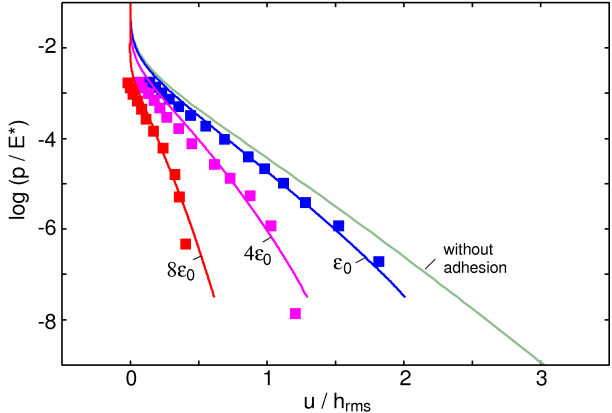}
\caption{\label{separation1}
Molecular dynamics (square symbols) and theoretical (solid lines) 
results for the (natural) logarithm
of the squeezing pressure $p$ (in units of $E^*$) as a function
of the average interfacial separation $\bar u$ in units of the root-mean-square
roughness amplitude $h_{\rm rms}$. The theoretical results are obtained from Eq.~\ref{interfacial.separation}.} 
\end{figure}

The MD simulations (square symbols) are compared with the theory (solid lines) for the 
relative contact area $A/A_0$ as a function of the squeezing pressure $p$ (see Fig.~\ref{last}). 
In Fig.~\ref{separation1} we show the 
(natural) logarithm
of the squeezing pressure $p$ (in units of $E^*$) as a function
of the average interfacial separation $\bar u$ in units of the root-mean-square
roughness amplitude $h_{\rm rms}$. 
The theoretical results (solid lines) agree very well with the 
molecular dynamics calculations (square symbols). 

Benz et al.\cite{MB} have studied the contact mechanics for polymer surfaces. They studied the
interfacial separation as a function of the squeezing pressure, and found an 
absolute (average) slope bigger
than the one obtained
from Finite Element calculations\cite{Rob} for non-adhesive contact 
between the interfacial separation and the log-scale pressure.
For adhesive contact, both theory and MD simulations predict a 
bigger absolute (average) slope
than the one for non-adhesive contact (see Fig.~\ref{separation1}).
However for the system studied by Benz et al.\cite{MB} we find that the adhesional interaction
gives only a very small ($\approx 20 \%$) increase in the absolute (average) slope,  
which cannot explain the experimental results by Benz et al.\cite{MB}.
In the supplementary material\cite{supplement} we present a critical analysis of this point.

To summarize, we have presented a molecular dynamics (MD) 
study of the adhesive contact between elastic solids with randomly
rough surfaces. 
We have calculated the contact area and the interfacial separation between the elastic solids,
and compared the results with the predictions of a recently developed contact mechanics model, which is based
on continuum mechanics. Considering the uncertainty in how to define the contact 
area and the interfacial separation at the atomistic level, and the small size of 
our MD system, the agreement between the theory and the MD results is very good.

We thank U. Tartaglino for many useful discussions.

\end{document}